\documentclass[reprint,aps,secnumarabic,balancelastpage,amsmath,amssymb,floatfix,superscriptaddress,nofootinbib]{revtex4-1}
\usepackage[latin1]{inputenc}
\usepackage{amsmath}
\usepackage{amsfonts}
\usepackage{amssymb}
\usepackage[title]{appendix}
\usepackage[pdftex]{graphicx}

\usepackage{hyperref}
\usepackage[normalem]{ulem}
\hypersetup{
	colorlinks = true,
	citecolor = {red}
	}

\begin{document}

\title{\bf Machine learning, quantum chaos, and pseudorandom evolution}

\author{Daniel W.F. Alves}
\email{dwfalves@gmail.com}
\affiliation{\footnotesize Center for Quantum Mathematics and Physics, Department of Physics, University of California, Davis \break Davis, CA, 95616, USA}
\affiliation{\footnotesize UNESP - Sao Paulo State University, Institute for Theoretical Physics (IFT) \break R. Dr. Bento T. Ferraz 271, Bl. II, 01140-070, Sao Paulo, SP, Brazil}
\author{Michael O. Flynn}
\email{miflynn@ucdavis.edu}
\affiliation{\footnotesize Center for Quantum Mathematics and Physics, Department of Physics, University of California, Davis \break Davis, CA, 95616, USA}

\date{\today}

\begin{abstract}

By modeling quantum chaotic dynamics with ensembles of random operators, we explore how machine learning learning algorithms can be used to detect pseudorandom behavior in qubit systems. We analyze samples consisting of pieces of correlation functions and find that machine learning algorithms are capable of determining the degree of pseudorandomness which a system is subject to in a precise sense. This is done without computing any correlators explicitly. Interestingly, even samples drawn from two-point functions are found to be sufficient to solve this classification problem. This presents the possibility of using deep learning algorithms to explore late time behavior in chaotic quantum systems which have been inaccessible to simulation. 




\end{abstract}

\maketitle

\section{Introduction}

Exploring the real-time dynamics of quantum systems is a central problem in modern theoretical physics. On the one hand, recent experiments in cold-atom physics \cite{2016PhRvA..94d0302S,2014Natur.507..475F} have allowed physicists to study the dynamics of isolated quantum systems. This is essential for deepening our understanding of both quantum mechanics and statistical mechanics \cite{2008Natur.452..854R,PhysRevA.43.2046,1994PhRvE..50..888S}. On the other hand, the AdS/CFT correspondence has linked the thermalization of isolated quantum systems to black-hole formation in dual higher-dimensional theories \cite{veronikareview}.

However, simulating the time evolution of even simple many-body lattice models has proven problematic. This difficulty has two sources: first, the dimensionality of Hilbert space scales exponentially with system size. Second, the sign problem is typically present in probability distributions associated with real-time evolution \cite{2008PhRvL.100q6403M,2009PhRvB..79c5320W}. These problems limit the effectiveness of both exact methods and various numerical methods, particularly those in the Monte Carlo family. Other approaches, notably those involving tensor networks, are capable of circumventing many of these difficulties \cite{2014AnPhy.349..117O,2007PhRvL..99l0601L,2008PhRvL.101k0501V,2015PhRvL.115t0401E,2009PhRvB..80o5131G,realtime1,realtime2,realtime3}. In general, however, tensor network methods break down in the presence of massive entanglement - preventing the study of interesting phenomena such as scrambling \cite{2007PhRvL..98g0201V,2018arXiv180200801X}.
%

A central set of objects in the study of scrambling, and quantum chaos in general, are out-of-time-order correlation functions (OTOCs) \cite{Maldacena:2015waa}. In particular, the decay of 4-point OTOCs is a common diagnostic of scrambling. To understand this, consider a local unitary operator $B$ acting on a qubit system. Under time evolution, $B$ becomes a sum of nested commutators,

\begin{equation}
B\left(t\right)=e^{iHt}Be^{-iHt}=\sum_{n=0}^{\infty}\frac{\left(it\right)^{n}}{n!}[H,...\left[H,B\right]...]
\end{equation}

For a chaotic system (alternatively, a sufficiently ``generic'' Hamiltonian \cite{2011PhRvL.106e0405B}), this will be a highly complicated, non-local operator. This growth of local operators can be quantitatively probed using a second local unitary operator, $A$, by considering an OTOC \cite{2018arXiv180200801X}:

\begin{equation}\label{thermaltrace}
\langle A^{\dagger}B^{\dagger}(t)AB(t)\rangle_{\beta}\equiv\langle A^{\dagger}U^{\dagger}B^{\dagger}UAU^{\dagger}BU\rangle_{\beta}
\end{equation}

where $\langle\cdot\rangle_{\beta}$ denotes a thermal expectation value in a state with temperature $\beta^{-1}$ and $U$ is a time evolution operator. For simplicity, and following most of the literature, we will consider the infinite temperature expectation value, i.e.,

\begin{equation}\label{thermaltrace2}
\langle A^{\dagger}B^{\dagger}(t)AB(t)\rangle_{\beta = 0}=\text{Tr}\left(A^{\dagger}B^{\dagger}\left(t\right)AB\left(t\right)\right)
\end{equation}

Assuming their support is disjoint, at early times $B(t)$ commutes with $A$, and the OTOC is $\mathcal{O}\left(1\right)$. As time passes, $B$ becomes increasingly non-local, the commutator $\left[B(t),A\right]$ grows, and the OTOC decays in most states. The time scale when this decay begins for 4-point functions is known as the scrambling time, $t_{\text{scr}}$ (analogous timescales are conjectured to exist for higher-point OTOCs  \cite{2018PhRvL.120l1601H,2017JHEP...04..121R}). This scrambling behavior implies the rapid growth of entanglement, so that computing OTOCs poses a serious challenge for reasons already discussed \cite{2018PhRvX...8b1013V,Nahum:2016muy,2016JHEP...02..004H,entanglementdynamics1,entanglementdynamics2,entanglementdynamics3,entanglementdynamics4}.

In the absence of a technique to compute OTOCs directly for generic lattice systems, it is interesting to ask if it is possible to detect scrambling without actually computing an OTOC. As a step towards answering this question, we consider ensembles of unitary operators known as $k$-designs. These ensembles form a hierarchy of pseudorandom operators, which have been used to model distinct regimes of chaotic quantum evolution (as reviewed in section \ref{II}). For instance, 1-designs cannot model scrambling, while $k$-designs with $k\geq 2$ can.

In this work, we demonstrate that it is possible to detect pseudorandomness associated with scrambling, without explicitly computing any thermal correlation functions, e.g., (\ref{thermaltrace2}). We accomplish this task by proposing a novel diagnostic of pseudorandomness which involves computing only a small number of terms in a given thermal correlation function, and presenting this data to a machine learning algorithm. The algorithm is then used to detect different levels of pseudorandomness in distinct correlation functions, which effectively diagnoses scrambling in our toy model. We have tested many classes of machine learning algorithms, but as we will show, only feedforward and convolutional neural networks were able to achieve reliable degrees of accuracy.

 We find for a variety of correlation functions that these machine learning architectures can reliably distinguish between three different ensembles of unitary operators (1-, 2-, and $\infty$-designs). Remarkably, we show that even terms taken from 2-point functions can be used as inputs for this classification problem. This constitutes a proof of concept that machine learning algorithms may be used to find interesting properties of quantum chaotic systems, such as time scales beyond and including the scrambling time, where other known methods tend to fail.

The rest of the paper is organized as follows. In section \ref{II}, we introduce some essential background on the phenomenology of chaotic quantum systems. We define $k$-designs and explain how they have been used to model different chaotic regimes. Section \ref{III} provides a brief summary of the machine learning architectures used to classify our data. Finally, our machine learning results are presented and discussed in  section \ref{IV}, with a summary and additional commentary in section \ref{V}.

\section{Chaos and Unitary Designs}\label{II}

In recent years, it has become clear that some physical systems can be modeled efficiently with random dynamics \cite{Nahum:2016muy,2018PhRvX...8b1014N,2018PhRvX...8b1014N,random1,Hayden:2007cs}. Physically, this is justified for systems which rapidly scramble information, and indeed it has been found that black holes are optimally fast scramblers in a certain sense \cite{Maldacena:2015waa,2008JHEP...10..065S}. Mathematically, such a distribution is known as the Circular Unitary Ensemble (CUE), and it is designed to replicate the Haar measure on the unitary group \cite{Guhr:1997ve}.

While it is often a dramatic simplification to replace physical time-evolution operators with random matrices, sampling from the CUE is often unnecessary. In practice, it is usually sufficient to sample from an ensemble of operators which reproduce the first $k$ moments of the CUE; such an ensemble is called a $k$-design. 

More formally, a $k$-design ${\cal E}=\left\{ p_{j},U_{j}\right\}$ consists of a set of unitary operators $U_{j}$, each associated with a probability $p_{j}$, which satisfies

\begin{equation}
\sum_{j}p_{j}U_{j}^{\otimes k}\rho\left(U_{j}^{\dagger}\right)^{\otimes k}=\int_{\text{Haar}}dU\,\, U^{\otimes k}\rho\left(U^{\dagger}\right)^{\otimes k}
\end{equation}

for any quantum state $\rho$. Note that a $k$-design automatically constitutes a $(k-1)$-design, and that the CUE itself  is an ``$\infty$-design''.

\begin{figure}
\hspace{-24pt}
\includegraphics[width=0.52\textwidth]{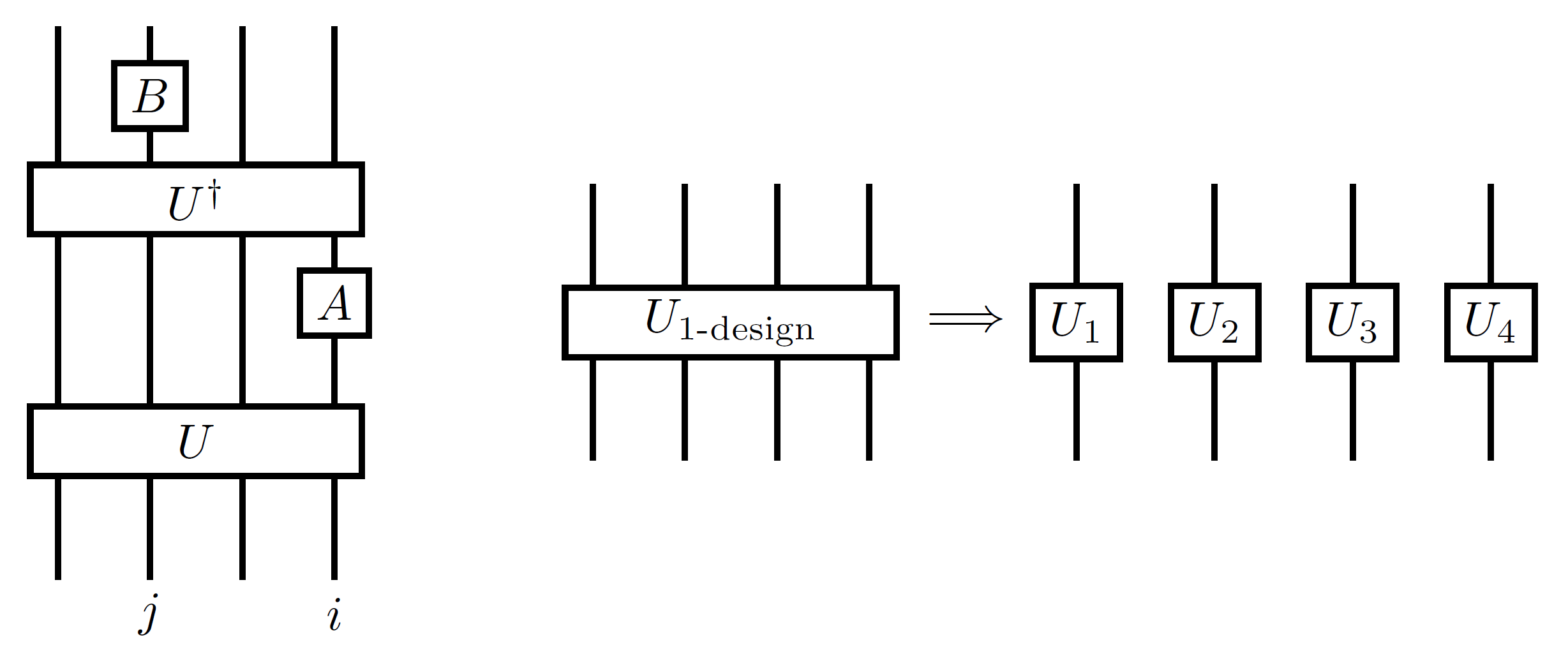}\caption{Left: a circuit diagram for the two-point function sample $\langle\sigma|B_{j}U^{\dag}A_{i}U|\sigma\rangle$. Right: An illustration of the factorization of 1-designs into single-site operators: this prevents them from scrambling information. Higher designs cannot be factorized, accounting for their ability to model scrambling.}
\label{fig:circ}
\end{figure}

For an arbitrary $k$-design, there is no unique or canonical way to generate members of the ensemble. For 1-design representatives, we will randomly draw elements of the Pauli group \cite{2017JHEP...04..121R}, and we will construct 2-design elements\footnote{This prescription only constructs $\epsilon$-approximate 2-designs. Throughout we will take $\epsilon=10^{-3}$ using the norm discussed in \cite{2017JMP....58e2203N}.} according to the procedure in \cite{2017JMP....58e2203N}. Physically, 1-designs are unique among designs because they cannot be used to model scrambling. This is because they can be factorized into operators acting on single qubits (see figure \ref{fig:circ}).

It is straightforward to see why $k$-designs are desirable for use in practice: sampling an operator from the CUE faces an exponential cost $\mathcal{O}\left(e^{N}\right)$. There are, however, circuits which can construct particular $k$-designs that grow only as $\mathcal{O}(\text{poly}(N))$ \cite{2017JMP....58e2203N,2010arXiv1010.3654B}. Further, $k$-designs have found other distinct applications in quantum information theory \cite{Adam2013thesis}.

In particular, chaotic dynamics have been modeled with $k$-designs for the purpose of computing correlation functions. To understand this in context, let us briefly review some aspects of real-time dynamics for strongly coupled thermal systems. For 2-point functions, one typically finds

\begin{equation}
\langle A B\left(t\right)\rangle\rightarrow\langle A\rangle\langle B\rangle+\mathcal{O}\left(e^{-t/t_{\text{th}}}\right),\,\,\,\,\,\, t>t_{\text{th}},
\end{equation}

where $t_{\text{th}}$ is the thermalization time. For OTOCs, a similar result is anticipated between the thermalization time and the scrambling time:


\begin{equation}
\langle AB\left(t\right)CD\left(t\right)\rangle\rightarrow\langle AC\rangle\langle BD\rangle+\mathcal{O}\left(e^{-t/t_{\text{th}}}\right), 
\end{equation}

$t_{\text{th}}<t<t_{\text{scr}}$. However, beyond the scrambling time, the OTOC further decays exponentially to a small value: 

\begin{equation}
\langle AB\left(t\right)CD\left(t\right)\rangle\rightarrow\mathcal{O}\left(\epsilon\right),\,\,\,\epsilon<<1,\,\,\,\,\,\,\,\,\,t>t_{\text{scr}}
\end{equation}

 More details on the scaling of this and related correlators can be found in \cite{2017JHEP...11..048C}. As shown in \cite{2017JHEP...04..121R}, $k$-designs can be used to model these results. For instance, substituting the time evolution operator with an average over a 1-design ensemble in the 2-point function gives


\begin{equation}
\sum_{j}p_{j}\langle AU_{j}^{\dagger}BU_{j}\rangle \rightarrow \langle A\rangle\langle B\rangle,
\end{equation}

where $\{p_{j},U_{j}\}$ is a 1-design. Moreover, any $k$-design with $k\geq 1$ suffices to guarantee this behavior. Further, a 4-point OTOC with (non-identity) Pauli operators $A,B,C,$ and $D$ satisfies

\begin{equation}
\sum_{j}p_{j}\langle AU_{j}^{\dagger}BU_{j}CU_{j}^{\dagger}DU_{j}\rangle\rightarrow \mathcal{O}\left(e^{-N}\right),
\end{equation}

whenever $\{U_{j},p_{j}\}$ forms a $k$-design with $k\geq 2$. Together, these results suggest that 1-designs can capture some important aspects of the physics between the thermalization and scrambling times. 2-designs capture comparable details after the scrambling time (until some other time scale). Later time scales, corresponding to the decay of higher-point OTOCs, are conjectured to exist. Between these time scales, it is believed that chaotic systems are effectively modeled by appropriately chosen designs \cite{2017JHEP...04..121R}.

Overall, this suggests that the dynamics of chaotic systems can be modeled by operators of increasing pseudorandomness as time grows. In some situations, the AdS/CFT correspondence has been used to find this hierarchy of timescales and OTOC decays explicitly \cite{2018PhRvL.120l1601H}.

Taking these results into account, we will now use different $k$-design ensembles as toy models for the degrees of pseudorandomness that a time evolution operator achieves during physical evolution. We will use a machine learning algorithm to analyze samples of correlation functions as input data, when computed for different $k$-designs, and see that we can reliably determine the value of $k$. We will comment on the possible generalization of these results to physical time-evolution operators in section \ref{V}.

\section{Supervised machine learning}\label{III}

The field of artificial intelligence (AI) has gained enormous momentum in recent years due to advances in algorithm efficiency and hardware power. In recent years, AI has been used to differentiate phases of matter \cite{2017NatPh..13..431C,2017PhRvE..95f2122H}, accelerate Monte Carlo simulations \cite{2017PhRvB..95c5105H}, and develop both variational and exact representations of wave functions \cite{2017Sci...355..602C,2018arXiv180209558C,2017PhRvX...7b1021D,2017PhRvB..96s5145D}. In the next section, we will show how to map the problem of determining the level of pseudorandomness characteristic of a system's evolution onto an image recognition problem. The task of classifying images is ideally suited to a class of deep learning algorithms known as Convolutional Neural Networks (CNNs), although as we will see, simpler architectures (feedforward neural networks) also succeed in the problem of interest. In this section, we will briefly comment on how a machine is trained to detect features in images.

To start the learning process, we consider a dataset consisting of several thousand images. Each image carries a label which can take one of two values; in our case, the label corresponds to the value $k$ of the $k$-design that was used to generate the image. We randomly separate these labeled images into two sets, a training set and a validation set. The machine learning algorithm analyses the images in the training set and their corresponding labels, looks for statistical patterns in the distribution of pixels, and automatically searches for features which allow it to distinguish between the two image classes with high accuracy. This is accomplished in the following a way: given the training set as an input, the algorithm applies to them a non-linear function that has hundreds of free parameters which are initialized randomly. The output of that function is either zero or one, corresponding to the labels associated with the images. We then define a cost function that quantifies how far the result of this operation is from the labels that accompanied the inputs. The free parameters are then adjusted by an appropriate optimization algorithm, such as gradient descent, in such a way as to minimize this cost function. By repeating this process, a minimum of the cost function is eventually found.

Afterwards, we validate our findings by asking the machine to classify images in the validation set. This set consists of thousands of images which the machine did not see in the training phase. This machine learning scheme, in which the inputs to the machine are accompanied by their labels, is known as ''supervised learning''. We refer the reader to \cite{hastie01statisticallearning} for more details on the supervised learning approach.

\begin{figure}
 \centering
 \includegraphics[width=0.40\textwidth]{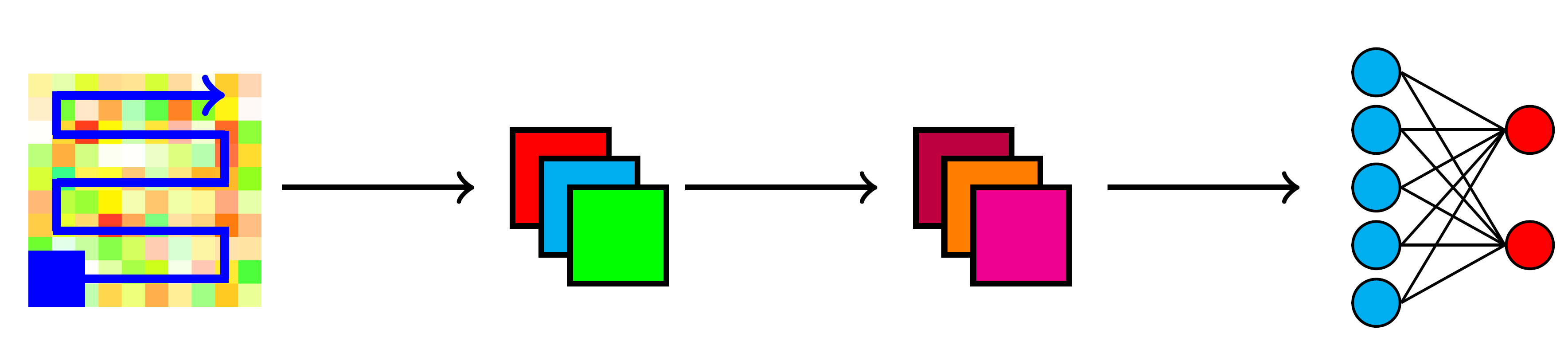}
 \caption{A schematic diagram of the convolutional neural network architecture used in this work. The blue square represents a filter. It takes the pixels covered by it and applies a non-linear function to them, outputting a single new pixel. We then slide the filter over the whole image, which produces a new image as its final output. We repeat this for each of the filters. This set of operations constitutes one convolutional layer. This is followed by a second convolutional layer. Next, the output is flattened and run through a fully connected layer. After applying a softmax function, the outputs are between 0 and 1, corresponding to probabilities that the given image is in a particular class. In a non-convolutional neural network, the input matrix is simply flattened and goes to the fully connected layer. }
 \label{cnnfigure}
 \end{figure}

Next, we will briefly outline the algorithm and training procedure for neural networks. We should note that other machine learning algorithms, such as random forest and XGBoost were also implemented, but the accuracy achieved in those cases was never significantly higher than 0.5. To start the learning procedure, one parses input images as sets of real valued matrices. In the convolutional neural network construction, these matrices are then convolved with a set of filters, which work as follows (see figure \ref{cnnfigure}).

 The blue square represents a filter that is applied to a set of pixels in the image. First, the pixels covered by the filter are ``flattened'' from a matrix into a vector $X_{i}$. We then apply the following linear transformation on these pixels, yielding a single new pixel: 

 \begin{equation}\label{flatten}
 X_{i} \rightarrow W_{i,j}X_{i}+b_{j}
 \end{equation}

 Where $W$ and $b$ are known as ``weights'' and ``biases'', which are free parameters that are optimized during the learning phase. We may have more than one filter, and each filter has a corresponding set of weights and biases. 

 We then introduce non-linearities by applying a non-linear function to this output, usually the so-called rectified linear unit, which we denote by $\sigma$:

 \begin{equation}\label{nonlinear}
 \sigma\left(x\right)=\text{max}\left(0,x\right)
 \end{equation}

 In a convolutional neural network, we ``slide'' the filters over the entire image as explained in figure \ref{cnnfigure}, performing the convolution operation for each set of pixels covered by the filter. The output after this convolution step is a set of new images, one for each filter.  

 In general, a pooling layer is then used which replaces subsets of the data by their maximal values. In our case, this is not used, as the size of the images in our dataset is reasonably small. These convolution steps may, in principle, be repeated dozens of times for a single input. 

 Next, the output is flattened and run through a fully-connected layer. In this step, the images of the preceding convolution step are flattened into a vector. Then, we consider all possible pairings of pixels in the flattened image with neurons, represented in figure \ref{cnnfigure} as blue circles. For each such connection, we apply the operation (\ref{flatten}) and the non-linearity (\ref{nonlinear}) again. Here, $(i,j)$ refers to a pair (pixel,neuron), so that each such connection has its own weight and each neuron has its own bias. Finally, another fully connected layer follows, but with only two outputs (figure \ref{cnnfigure}). After applying a softmax function to these outputs, their values sum up to one. We can then interpret these values as probabilities, corresponding to the machine's certainty that the provided input is in a particular class of images.

In a non-convolutional, feedforward neural network, we simply flatten the input matrix and go straight to the fully connected layer in the setup described above.

 We then define a cost function, $C$, to quantify how far the outputs of the machine are from the labels that accompany the inputs. Since the output depends on the weights and biases, $C$ must depend on them. In principle, the cost should be a sum over the costs that one gets after checking the outputs of every single image in the training set against their respective labels, 

 \begin{equation}
 C_{\text{total}}\left(W,b\right)=\sum_{i\in \text{training set}}C_{i}\left(W,b\right)
 \end{equation}

 where $C_{i}$ is the cost of a single input image. However, since the training set is usually too large, we define the cost over randomly chosen mini-batches of the training set and run gradient descent (see below) on each of them.

 At each training step, the machine takes the images in the batch as input, runs them through the various layers, and updates the values of the weights and biases  using a gradient descent algorithm:

 \begin{equation}
 W^{t+1}=W^{t}-\eta\nabla C\left(W,b\right)
 \end{equation}

 \begin{equation}
 b^{t+1}=b^{t}-\eta\nabla C\left(W,b\right)
 \end{equation}

 Where $\left(W^{t},b^{t}\right)$ are the set of weight and biases at step $t$ and $\eta$ is a free parameter chosen by hand, known as the learning rate. In this work, we used the Adam algorithm \cite{ADAM}, where different learning rates are associated to different weights and biases of the network, and are adapted during the training. After this step is completed, we draw another batch from the training step and repeat the process. Letting $n_{\text{batch}}$ be the size of the batch and $n_{\text{training}}$ be the total number of images in the training set, we say we have completed one epoch of training after repeating the whole process $n_{\text{training}}/n_{\text{batch}}$ times. We then train the network for as many epochs as desired. To validate the model, we apply the neural network to batches of images in the validation set, and compute the accuracy obtained.  

The cost function employed in this work is the cross-entropy, which is a standard choice for image recognition applications. The cross-entropy is also convenient because it speeds up the learning process when compared to a naive mean-squared error. Again, we refer the reader to \cite{nielsenneural} for more details. 

 Some parameters like the learning rate, the batch-size, the size of the filters and the way they are initialized, do not change during the learning process and must be adjusted by hand. These are known as ``hyper-parameters''. 

 The convolutional neural network architecture has been extremely successful for image recognition problems, due to the fact that the convolutions are able to learn local features across the image. The non-convolutional architecture can also be used for image recognition tasks, but is usually less expressive when applied to natural image datasets. We have experimented with CNNs and feed-forward networks; the results obtained in both cases were very similar. However, for the sake of coherence in our presentation, we only discuss our results from the perspective of CNNs. The comparable success of the feed-forward structure can be attributed to the lack of highly detailed features in our datasets, unlike the varied real-world images used to train machines for other tasks.
 
 These algorithms have been applied to many-body physics to detect phase transitions in \cite{2017NatPh..13..431C,2017NatSR...7.8823B}, using Monte Carlo simulation configurations as input data. Though there are universality theorems that prove that any function can be approximated with a neural network ansatz, it is still unclear why the number of neurons and filters necessary does not scale exponentially with the size of the input for most problems. There are proposals for future work which address this question \cite{2017JSP...168.1223L}.
 
 \section{Visualizing Correlation Functions}\label{IV}

We turn now to the task of mapping samples of correlation functions onto images and analyzing them with the CNN. This process is inspired by the work of Broecker et. al. \cite{2017NatSR...7.8823B}, where a similar technique was applied to study many-body fermion systems with a sign problem. Their goal was to find a way to differentiate between phases in such systems while avoiding the computational issues associated with the sign problem. This was accomplished by computing samples of an order parameter (in their case, a Green's function) and mapping the resulting samples onto images. Machine learning methods similar to those presented here were then successful in differentiating these images.

Here, we can regard the different regimes of chaotic evolution (corresponding to models described by different $k$-designs) as distinct ``phases''. Our ``order parameters'' are the values of OTOCs. Since these OTOCs cannot be efficiently computed in general, we compute only small samples of them (see (\ref{sample}) for details). For example, consider the ensemble-averaged 4-point function

\begin{equation}\label{fullsum}
\sum_{n}p_{n}\langle A_{i}^{\dagger}U^{\dagger}_{n}B_{j}U_{n}C_{i}U^{\dagger}_{n}D_{j}U_{n}\rangle_{\beta=0}
\end{equation}

where $i$ and $j$ index qubit sites, $U_{n}$ is drawn from a $k$-design with probability $p_{n}$, and $A, B, C$ and $D$ are taken to be (non-identity) Pauli matrices. Given that the trace over qubit states is difficult to compute, we consider a piece of the sum (\ref{fullsum}) which amounts to a random sample of the correlator:

\begin{equation}\label{sample}
\sum_{n=1}^{m}\langle\sigma_{n}|A_{i}^{\dagger}U_{n}^{\dagger}B_{j}U_{n}C_{i}U_{n}^{\dagger}D_{j}U_{n}|\sigma_{n}\rangle
\end{equation}

The ``batch number'' $m$ is chosen to be of order one to render the sample computable. We will take $m=5$ throughout. Each state $|\sigma\rangle$ is chosen randomly, and each unitary $U$ is drawn independently according to the appropriate ensemble. Calculating this sample of the correlation function yields a matrix of complex numbers corresponding to all possible locations of the operator insertions $(i,j)$. Each of these numbers can be mapped onto a colored pixel using the HSV color scheme \cite{Foley:1990:CGP:83821}, where the amplitude of the complex number determines the hue and the phase sets the saturation.

We then consider a pair of correlators with identical operator insertions but with unitaries drawn from different $k$-designs. For each of these correlators, we generate 3125 images for a total of 6250. The training set contains 5000 randomly chosen images from this set, so that the validation set contains 1250. These images are then given to the CNN, and the machine attempts to classify the images by the ensemble which generated them.

We have examined a number of correlators, including both 2-point functions and 4-point OTOCs. Specifically, we started by generating images from 

\begin{equation}\label{xyxy}
\langle X_{i}U^{\dagger}Y_{j}UX_{i}U^{\dagger}Y_{j}U\rangle
\end{equation}

for both 2-designs and Haar-random matrices. If we were to use 1-designs for this OTOC, the result would be trivial, since 1-design operators factor into a product of single site Pauli matrices (see fig. \ref{fig:circ}). Hence the correlator can only take two distinct value (one for $i=j$ and another for $i\neq j$). To compare 1- and 2-designs in a non-trivial way, we also considered  images generated from

\begin{equation}
\langle X_{i}U^{\dagger}X_{j}UY_{i}U^{\dagger}Y_{j}U\rangle
\end{equation}

Besides these OTOCs, we also considered the following 2-point functions:

\begin{equation}
\langle X_{i}U^{\dagger}Y_{j}U\rangle
\end{equation}

\begin{equation}\label{zz}
\langle Z_{i}U^{\dagger}Z_{j}U\rangle
\end{equation}

These objects are not diagnostics of chaos. In particular, the ensemble-averaged value of these correlators will be the same for any $k$-design - including 1-designs (which cannot model scrambling). Nevertheless, samples of these correlators may vary as we change the design from which $U$ is sampled, and we can ask if the CNN is able to detect a difference between 1- and 2-designs.

We have restricted our system size to $N=10$ qubits in order to efficiently sample from the CUE (an ``$\infty$-design'') in addition to 1- and 2-designs. Earlier data sets considered smaller system sizes ($N=7$); these were found to have insufficient information for the CNN to find any distinction between the ensembles.

\begin{figure}
\includegraphics[width=0.53\textwidth]{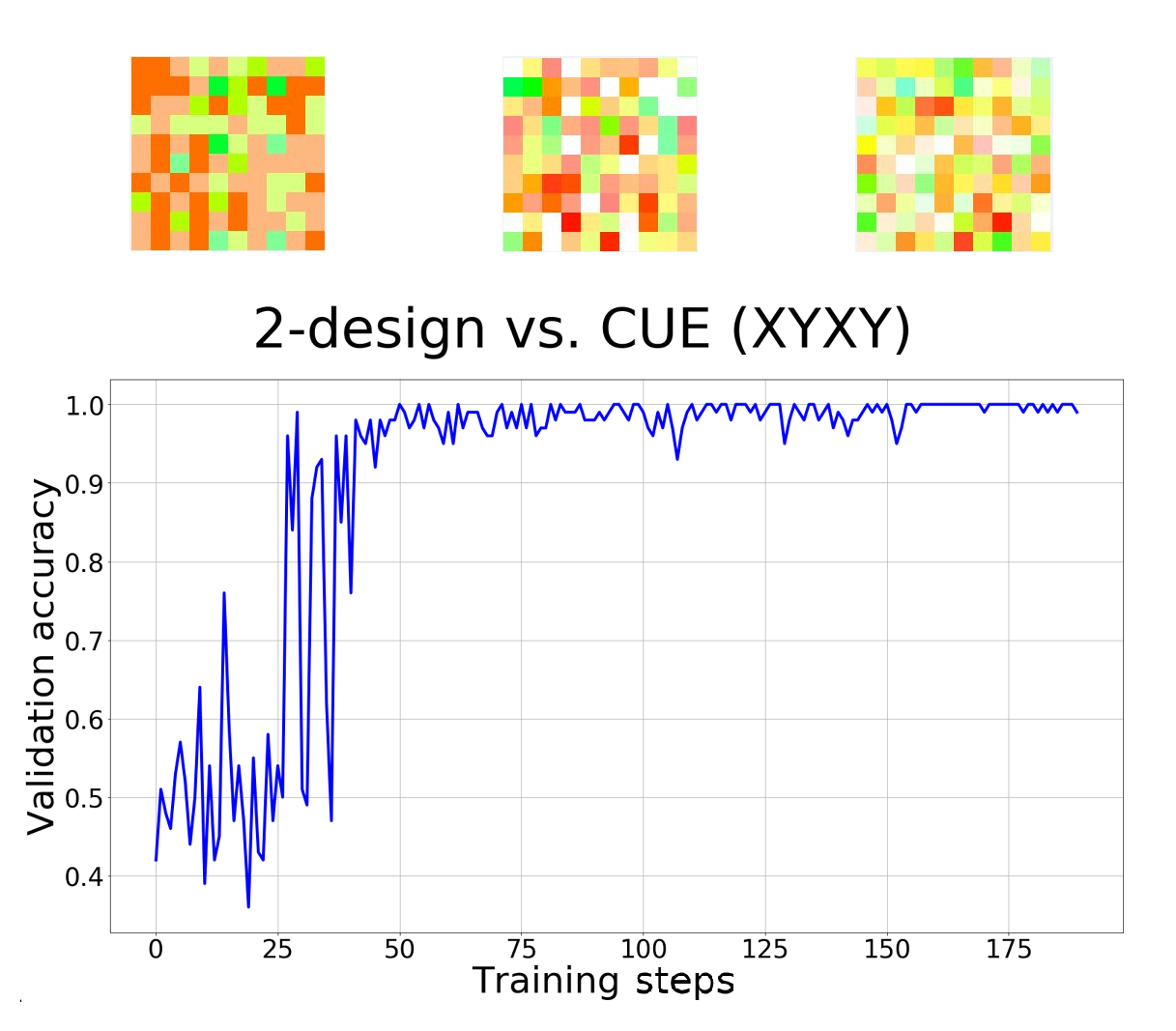}
\caption{Above (left to right): samples of image configurations for 1-designs, 2-designs, and Haar-random matrices generated from the correlator (\ref{xyxy}). With some effort, the human eye is capable of distinguishing 1-designs from the other two cases, while that seems quite challenging for the 2-design and Haar data. Below: Sample training outputs for the case of 2-designs vs. Haar matrices as a function of the number of training steps. Clearly, the machine is capable of distinguishing the two cases reliably given sufficient training. The training process takes several minutes on a typical laptop CPU.}
\label{dataone}
\end{figure}

In every case studied, our neural network is capable of  distinguishing each of the different types of design remarkably well, with over 99\% accuracy, as illustrated in figure \ref{dataone}. This is achieved after roughly 100 training steps, which takes several minutes on a modest laptop CPU. Though it is in general hard for the human eye, the CNN can identify some feature of each set of images which is unique to it. Hence, the CNN is able to detect pseudorandomness associated with scrambling without explicitly computing a thermal correlation function. This is true for each of the four correlation functions considered. Importantly, for a given correlator, we have checked that the machine is not capable of distinguishing between random samples of the same type of design: that is, the machine cannot distinguish $k$-designs from other $k$-designs. This strongly suggests that the feature which the CNN is detecting can only be accounted for by the level of pseudorandomness present in the unitaries used to compute the ensemble average.

The fact that the CNN can distinguish between each of these ensembles for every correlator considered is somewhat surprising. Given that a 1-design is sufficient to cause 2-point functions to decay, one would not expect to be able to use 2-point functions to distinguish between 1-designs and higher-order ensembles. Similarly, 4-point OTOCs seem insufficient to distinguish 2-designs from CUE matrices. The fact that these cases are distinguishable strongly suggests that these samples are sensitive to the structure of the operators under study in a non-trivial way. While different designs may lead to the same value for particular correlators, we believe that the means by which they go about canceling different contributions to the correlator contains additional structure. This allows us to detect higher-design-like behavior with particularly simple correlation functions.

\section{Discussion \& Conclusions}\label{V}

In this paper, we have put forward the idea of using deep learning to probe the real-time evolution of chaotic quantum systems. As a first step in this program, we modeled time evolution with pseudorandom operators sampled from several distinct $k$-design ensembles. This was motivated by the results of \cite{2017JHEP...04..121R}, and the generic conjecture that chaotic time-evolution operators become increasingly pseudorandom over time. By training a convolutional neural network with data samples drawn from correlation functions, we have shown that it is possible to determine the level of pseudorandomness in a system's evolution by solving a simple image recognition problem. Importantly, this can be done with very modest computational resources. This allows us to detect pseudorandomness associated with scrambling behavior with none of the traditionally expensive computations associated with that task. We have also shown that even samples from 2-point functions (which are not OTOCs) encode enough information to detect pseudorandom behavior associated with scrambling.

There are several lines of inquiry worth pursuing in light of these results. First, it would be interesting to extend this thought process to systems that are chaotic but with a deterministic Hamiltonian. For instance, the Ising model with certain choices of parameters exhibits a range of thermalization behaviors \cite{2011PhRvL.106e0405B}. We are exploring this question now \cite{BLAH}. In that case, as time passes, the expectation is that the CNN can detect the exact time at which the classification changes, corresponding to the scrambling time. More speculatively, one might hope this could be extended to find other time scales of physical interest.

Another interesting future direction is in the usage of unsupervised learning, as opposed to the supervised learning scheme used here. In unsupervised learning, the inputs to the algorithm do not carry labels, and the machine has to create labels by itself. That may allow patterns to be recognized that go beyond those explored in this work.

It would also be interesting to check if samples of other observables, besides the correlators used here,  can serve as useful diagnostics of chaos. Since $k$-designs have been used to model the way black holes process information, it is interesting to consider any possible insight our results may bring to black-hole physics. Finally, as $k$-designs are used routinely in quantum information theory, it is also worth looking for possible experimental consequences of our results \cite{2018arXiv180709087V,2016PhRvA..94f2329Z,2016arXiv160701801Y,2017PhRvX...7c1011L,2017arXiv170506714M}.

\section*{Acknowledgements}
We would like to thank Mukund Rangamani, Rajiv R.P. Singh, Veronika Hubeny, Xiao-liang Qi, Tommaso Roscilde, Tiago Mendes, Jerome Dubail, Isaac Kim, Tomonori Ugajin, Wenjian Hu, Wei-Ting Chiu, Benjamin Cohen-Stead, and James Sully for useful discussions and support. D.W.F.A. would like to acknowledge the hospitality of the Galileo Galilei Institute for theoretical physics and INFN, at Florence, Italy and the the University of California, Davis where part of this work was developed. D.W.F.A is supported by the PDSE Program scholarship of CAPES-Brazilian Federal Agency for Support and Evaluation of Graduate Education within the Ministry of Education of Brazil, and by the CNPq grant 146086/2015-5.

\bibliography{correlationimages}

\end{document}